%% file: main.tex
\documentclass[PHM,2019]{PHMSociety}

\usepackage{graphicx}
\usepackage{amsmath}
\usepackage{amssymb}  

\usepackage{tikz}
\usepackage{verbatim}
\usetikzlibrary{arrows,shapes,fit,scopes,calc,matrix,positioning,decorations.pathmorphing,fadings}

\usepackage[justification=centering]{caption}

\newtheorem{example}{Example}[section]

\begin{document}

\title{Isolation and Localization of Unknown Faults Using Neural Network-Based Residuals}

\author{%
	Daniel Jung$^1$
}

\address{
	\affiliation{1}{Department of Electrical Engineering, Link\"{o}ping University, Link\"{o}ping, 581 83, Sweden}{ 
		{\email{daniel.jung@liu.se}}
		} 
}

\maketitle

\phmLicenseFootnote{Daniel Jung}

\begin{abstract}
Localization of unknown faults in industrial systems is a difficult task for 
data-driven diagnosis methods. The classification performance of many 
machine learning methods relies on the quality of training data. Unknown 
faults, for example faults not represented in training data, can be detected 
using, for example, anomaly classifiers. However, mapping these unknown faults 
to an actual location in the real system is a non-trivial problem. In model-based 
diagnosis, physical-based models are used to create residuals that isolate 
faults by mapping model equations to faulty system components. 
Developing sufficiently accurate physical-based models can be a time-consuming 
process. Hybrid modeling methods combining physical-based 
methods and machine learning is one solution to design data-driven 
residuals for fault isolation. In this 
work, a set of neural network-based residuals are designed by incorporating 
physical insights about the system behavior in the residual model 
structure. The residuals are trained using only fault-free data and 
a simulation case study shows that they can be used to perform fault isolation 
and localization of unknown faults in the system.
\end{abstract}

\input{introduction}
\input{casestudy}

\input{background}

\input{residual}
\input{evaluation}
\input{conclusions}

\bibliographystyle{apacite}
\bibliography{references}

\section*{Biographies}

\noindent\textbf{Daniel Jung} was born in Link\"{o}ping, Sweden in 1984. He 
received a Ph.D. degree in 2015 from Link\"{o}ping University, Sweden. During 2017, 
he was a Research Associate at the Center for Automotive Research at The 
Ohio State University, Columbus, OH, USA. Since 2018, he is an Assistant Professor 
at Link\"{o}ping University. His current research interests include theory and applications 
of model-based and data-driven fault diagnosis, smart grids, and optimal control of hybrid electric 
vehicles.

\end{document}

%% file: introduction.tex
\section{Introduction}
An important task of fault diagnosis of industrial systems is fault localization, i.e., 
identifying where faults are located in the system. 
Increasing system complexity and autonomous operation 
require that the system is reliable and able to detect faults early before any 
accidents or damages occur. Being able to identify a faulty component 
gives important information when deciding for suitable counter-measures to 
minimize costs and the risk of potential dangers. 

Machine Learning has been very successful in many applications, including 
image classification and text analysis. One example 
of such methods are neural networks and deep learning. However, 
some of the recent successes have been made possible thanks to the access 
to large amounts of training data \cite{jia2016deep}. In many fault diagnosis 
applications, collecting representative data is complicated and expensive, 
especially during the system development phase and early system life 
\cite{sankavaram2015incremental}. Even though incremental classification 
algorithms are able to improve performance over time, as more training data 
become available, it is still relevant to be able to identify likely fault locations 
of fault scenarios not covered in training data. This is important in, for example, 
troubleshooting \cite{pernestaal2012modeling}.  One solution to limited 
training data is to use physical-based models when implementing machine learning 
algorithms. 

The idea of using physical-based model structures in data-driven neural network 
design for fault diagnosis has been proposed in \cite{pulido2019state}. With 
respect to the mentioned work, this paper presents how to use neural network-based 
residuals and physical-based models to localize unknown faults in the system. In 
\cite{garcia2011improving} a model parameter estimation approach based on a 
partitioned system model is proposed.  

The benefits of combining model-based and data-driven fault diagnosis methods have 
also been discussed in \cite{tidriri2016bridging}. Hybrid diagnosis system designs, 
combining model-based residuals and machine learning classifiers, have been 
proposed in, for example, \cite{jung2017combined,tidriri2017generic,jung2018combining}. The methods in these 
mentioned papers, relies on residuals to perform fault isolation and classification. 
Therefore, residual generation is an important task during the diagnosis system 
design to achieve satisfactory fault isolation performance. With respect to these 
previous works, not only fault isolation is considered here but also localization of unknown 
faults.  

\subsection{Problem Statement}

Even though a data-driven classifier is able to identify when an unknown fault 
has occurred, it is non-trivial to localize the fault in the actual system without 
training data from that fault. One solution is to utilize physical insights about the 
system when implementing a machine learning algorithm. 

In this work, a simulation study is performed to investigate if it is possible to point out 
the fault location in a system using a set of unconventional neural network-based residuals
where the network design represents the structure of the system. The neural network 
design is implemented using Python~3 and PyTorch \cite{paszke2017automatic}.  
It is assumed that training data to train the neural network models are available 
from nominal system operation only, i.e. no data from any fault scenario are used during 
the training phase. The study shows that including some physical insights in the neural 
network design makes it possible to detect and localize system faults even though the 
networks are trained using data from nominal system operation only.        

For the neural network design, structural model decomposition methods are used 
on a structural representation of the system. A structural model describes the 
relationship between system variables without considering the analytical relation, 
i.e., it only describes which variables are included in each model equation \cite{blanke2006diagnosis}.
Different residual generation algorithms use the structural model to create 
computational graphs to evaluate the model equations from sensor data to 
compute a residual, see for example \cite{frisk2017toolbox,pulido2004possible}. 
If a detailed analytical model of the system is not available, a structural model 
representing the physical-based relations, describing the system behavior, can still 
be used to design neural-networks for residual generation.     

%% file: casestudy.tex
\section{A Non-Linear Two Tank Simulation Case Study}

To illustrate the proposed method, a non-linear two-tank simulation model 
is used to simulate sensor data. An illustration of the system is shown in 
Figure~\ref{fig:schematic} and the model dynamics are derived from the 
Bernoulli equation as
\begin{equation}
\begin{aligned}
\dot{x}_1 &= - d_1 \sqrt{x_1} + d_2 u,  \quad \dot{x}_2 = - d_3 \sqrt{x_1} + d_4 \sqrt{x_2}, \\ 
y_1 &= x_1, \quad y_2 = x_2, \quad y_3 = d_5 \sqrt{x_1}, \quad y_4 = d_6 \sqrt{x_2}
\end{aligned} 
\end{equation}
where $x_i$ is water level in tank $i$, $u$ is a known input flow in tank one, $y_1$ and 
$y_2$ measure the water level in each tank, respectively, $y_3$ and $y_4$ measure the 
out-flow from each tank, respectively, and $d_1, \ldots, d_6$ are model parameters.

\pgfmathsetmacro{\tanktwo}{-1.5}
\pgfmathsetmacro{\tap}{-0.2}
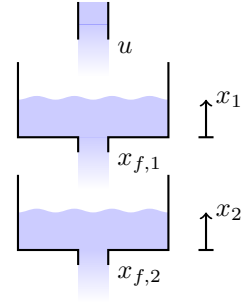
\begin{figure}[h]
\centering
      \begin{tikzpicture}
      
      	\path[fill=blue,opacity=0.2] (-0.2,2.0+\tap) -- (-0.2,1.7+\tap) -- (0.2,1.7+\tap) -- (0.2,2.0+\tap) -- cycle;
	\path[fill=blue,opacity=0.2,path fading = south] (-0.2,1.7+\tap) -- (-0.2,1.0+\tap) -- (0.2,1.0+\tap) -- (0.2,1.7+\tap) -- cycle;
        \draw[thick] (-0.2,2+\tap) -- (-0.2,1.5+\tap);
      	\draw[thick] (0.2,2+\tap) -- (0.2,1.5+\tap);
	
	\draw (0.2,1.4+\tap)  node[right]{$u$};
	
	\path[fill=blue,opacity=0.2] (-1,0) -- (-1,0.5) to[out=0, in=180] (-0.75, 0.55) to[out=0, in=180] (-0.5, 0.50) to[out=0, in=180] (-0.25, 0.55) to[out=0, in=180] (0, 0.50)
		     to[out=0, in=180] (0.25, 0.55) to[out=0, in=180] (0.50, 0.50) to[out=0, in=180] (0.75, 0.55) to[out=0, in=180] (1, 0.50) -- (1,0) -- cycle;
	
	\path[fill=blue,opacity=0.2,path fading = south] (-0.2,0) -- (-0.2,-0.7) -- (0.2, -0.7) -- (0.2, 0) -- cycle;
	
	\draw (0.2,-0.35)  node[right]{$x_{f,1}$};
	
	\draw[thick] (-1,1) -- (-1,0) -- (-0.2,0) -- (-0.2,0) -- (-0.2, -0.2);
	\draw[thick] (1,1) -- (1,0) -- (0.2,0) -- (0.2,0) -- (0.2, -0.2);
			     
	\draw[thick] (1.4, 0) -- (1.6, 0);
	\draw[thick, ->] (1.5, 0) -- (1.5, 0.5) node[right]{$x_1$};	

	\path[fill=blue,opacity=0.2] (-1,0+\tanktwo) -- (-1,0.5+\tanktwo) to[out=0, in=180] (-0.75, 0.55+\tanktwo) to[out=0, in=180] (-0.5, 0.50+\tanktwo) to[out=0, in=180] (-0.25, 0.55+\tanktwo) 
		to[out=0, in=180] (0, 0.50+\tanktwo) to[out=0, in=180] (0.25, 0.55+\tanktwo) to[out=0, in=180] (0.50, 0.50+\tanktwo) to[out=0, in=180] (0.75, 0.55+\tanktwo) 
		to[out=0, in=180] (1, 0.50+\tanktwo) -- (1,0+\tanktwo) -- cycle;
		     
	\path[fill=blue,opacity=0.2,path fading = south] (-0.2,0+\tanktwo) -- (-0.2,-0.7+\tanktwo) -- (0.2, -0.7+\tanktwo) -- (0.2, 0+\tanktwo) -- cycle;	     
		     
	\draw (0.2,-0.35+\tanktwo)  node[right]{$x_{f,2}$};	     
		     
	\draw[thick] (-1,1+\tanktwo) -- (-1,0+\tanktwo) -- (-0.2,0+\tanktwo) -- (-0.2,0+\tanktwo) -- (-0.2, -0.2+\tanktwo);
	\draw[thick] (1,1+\tanktwo) -- (1,0+\tanktwo) -- (0.2,0+\tanktwo) -- (0.2,0+\tanktwo) -- (0.2, -0.2+\tanktwo);

	\draw[thick] (1.4, 0+\tanktwo) -- (1.6, 0+\tanktwo);
	\draw[thick, ->] (1.5, 0+\tanktwo) -- (1.5, 0.5+\tanktwo) node[right]{$x_2$};	     
      \end{tikzpicture}
  \caption{An illustration of the two tank system.}
  \label{fig:schematic}
\end{figure}

In this case study, it is assumed that an accurate model of the system is 
not available for the diagnosis system design. Instead, a qualitative model is 
available that describes the general system behavior as follows: 
\begin{equation}
\begin{aligned}
e_1: & \dot{x}_1 = h_1(x_{f,1}, u) & e_5: & y_1 = x_1\\
e_2: & \dot{x}_2 = h_2(x_{f,1}, x_{f,2}) & e_6: & y_2 = x_2\\
e_3: & x_{f,1} = g_1(x_{1}) & e_7: & y_3 = x_{f,1}\\
e_4: & x_{f,2} = g_2(x_{2}) & e_8: & y_4 = x_{f,2}
\end{aligned}
\label{eq:watertank_model}
\end{equation}
where $x_{f,i}$ is out-flow from tank $i$, 
$u$ is a known input flow into tank one, and $y_1$, $y_2$, $y_3$, $y_4$ 
are sensor data. The functions $h_1(\cdot)$ and $h_2(\cdot)$ state that the change 
in water level in each tank depend on the inflow and outflow. The functions 
$g_1(\cdot)$ and $g_2(\cdot)$ say that the outflow depends on the water level in 
the tank. 

\begin{figure}[h]
      \begin{tikzpicture}
        \node[anchor=south west,inner sep=0] (image) at (0,0) {
          \resizebox{1\columnwidth}{!}{\includegraphics{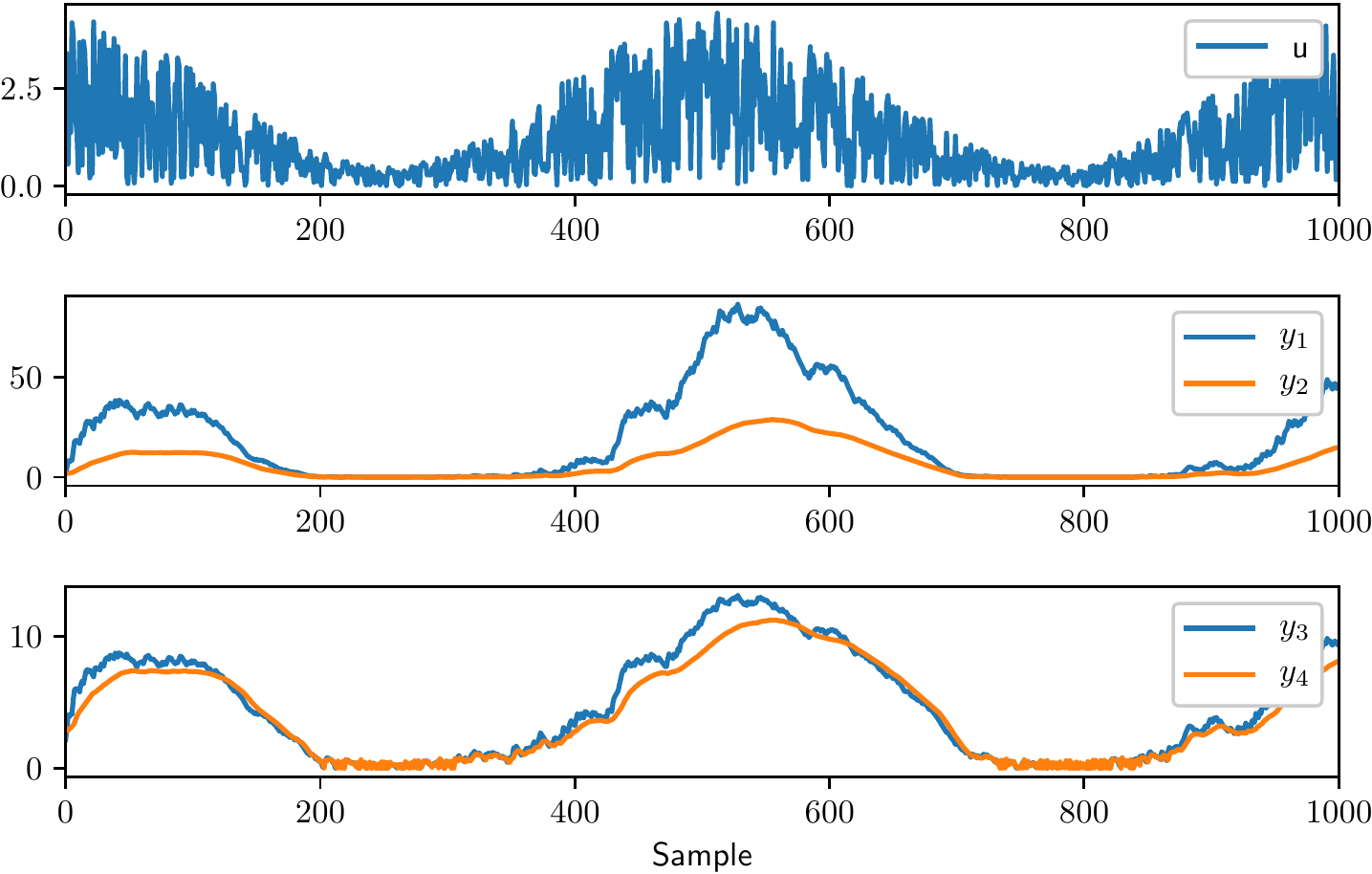}}
        };
        \begin{scope}[x={(image.south east)},y={(image.north west)}]
        \end{scope}
      \end{tikzpicture}
  \caption{An example of simulated data from the two tank system.}
  \label{fig:sim_data}
\end{figure}

An example of simulated data from the system is shown in 
Figure~\ref{fig:sim_data}. For the evaluation, the simulation model can be 
used to simulate different faults, for example leakages in the tank, clogging 
in the outflow pipes, and sensor faults.    

%% file: background.tex
\section{Background}

First, a brief summary of artificial neural networks is presented. Then, 
the principles of model-based diagnosis and structural analysis methods
are summarized.    

\subsection{Artificial Neural Networks}

Artificial neural networks and deep learning are a set of machine learning methods 
that can be used to approximate non-linear functions \cite{schmidhuber2015deep}. 
Neural networks consist of a set of neurons where the output from some neurons 
are inputs to other neurons and can be represented as a computational graph. 
Each neuron is a non-linear function of the inputs to the neuron, for example 
\begin{equation}
x_{i,\text{out}} = h_i(w_i^Tx_{i,\text{in}} + \beta_i)
\end{equation}
where $x_{i,\text{in}}$ denote the inputs to neuron $i$, $x_{i,\text{out}}$ the output, 
$w_i$ is a vector of weights, $\beta_i$ is a bias, and $h_i$ is a non-linear activation 
function, e.g. rectified linear unit (ReLU), sigmoid or hard tan \cite{aggarwal2018neural}. 
A common method for training neural networks is to use back-propagation. 

Conventional neural network designs arrange the neurons in different layers. The 
first layer of the neural network is denoted the input layer, the 
final layer is called the output layer, and all layers in between are called hidden layers. 
One type of neural networks, called recurrent neural networks, can be used to 
model temporal dynamic systems \cite{pearlmutter1995gradient}. 
In recurrent neural networks, the output from some neurons are used as input to 
other neurons at concurring time steps. This is shown in Figure~\ref{fig:neural_network}
where the state variable $\hat{x}_1$ is used as input in the next time instance of 
the recurrent neural network. The variable $u_t$ is an input signal and $\hat{y}_t$ is an 
output signal at time instance $t$.

\pgfmathsetmacro{\nn}{3.5}
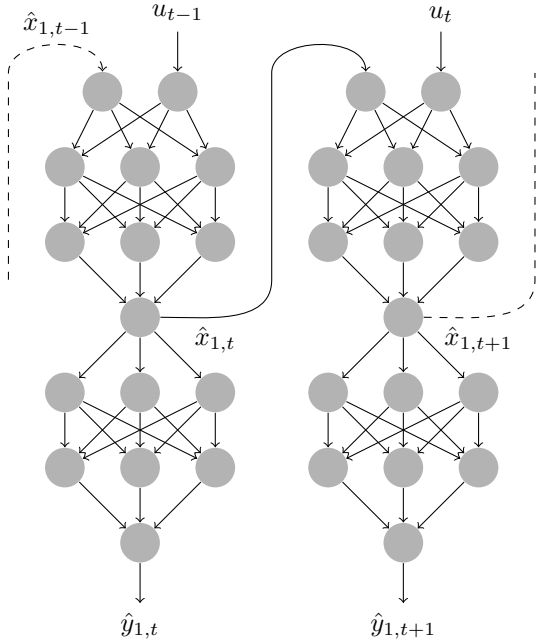
\begin{figure}[h]
\centering
\begin{tikzpicture}[]
    \tikzstyle{every pin edge}=[<-,shorten <=1pt]
    \tikzstyle{neuron}=[circle,fill=black!30,minimum size=15pt,inner sep=5pt]
    \tikzstyle{annot} = [text width=4em, text centered]

    \node[neuron] (N111) at (0,0) {};
    \node[neuron] (N112) at (1,0) {};
    \node[neuron] (N121) at (-0.5,-1) {};
    \node[neuron] (N122) at (0.5,-1) {};
    \node[neuron] (N123) at (1.5,-1) {};
    \node[neuron] (N131) at (-0.5,-2) {};
    \node[neuron] (N132) at (0.5,-2) {};
    \node[neuron] (N133) at (1.5,-2) {};
    \node[neuron] (N141) at (0.5,-3) {};
    \node[neuron] (N151) at (-0.5,-4) {};
    \node[neuron] (N152) at (0.5,-4) {};
    \node[neuron] (N153) at (1.5,-4) {};
    \node[neuron] (N161) at (-0.5,-5) {};
    \node[neuron] (N162) at (0.5,-5) {};
    \node[neuron] (N163) at (1.5,-5) {};    
    \node[neuron] (N171) at (0.5,-6) {};      
    
    \draw[->] (N111) -- (N121);\draw[->] (N111) -- (N122);\draw[->] (N111) -- (N123);
    \draw[->] (N112) -- (N121);\draw[->] (N112) -- (N122);\draw[->] (N112) -- (N123);
    \draw[->] (N121) -- (N131);\draw[->] (N121) -- (N132);\draw[->] (N121) -- (N133);
    \draw[->] (N122) -- (N131);\draw[->] (N122) -- (N132);\draw[->] (N122) -- (N133);
    \draw[->] (N123) -- (N131);\draw[->] (N123) -- (N132);\draw[->] (N123) -- (N133); 
    \draw[->] (N131) -- (N141);\draw[->] (N132) -- (N141);\draw[->] (N133) -- (N141); 
    \draw[->] (N141) -- (N151);\draw[->] (N141) -- (N152);\draw[->] (N141) -- (N153); 
    \draw[->] (N151) -- (N161);\draw[->] (N151) -- (N162);\draw[->] (N151) -- (N163); 
    \draw[->] (N152) -- (N161);\draw[->] (N152) -- (N162);\draw[->] (N152) -- (N163); 
    \draw[->] (N153) -- (N161);\draw[->] (N153) -- (N162);\draw[->] (N153) -- (N163); 
    \draw[->] (N161) -- (N171);\draw[->] (N162) -- (N171);\draw[->] (N163) -- (N171); 
    
    \node[neuron] (N211) at (0+\nn,0) {};
    \node[neuron] (N212) at (1+\nn,0) {};
    \node[neuron] (N221) at (-0.5+\nn,-1) {};
    \node[neuron] (N222) at (0.5+\nn,-1) {};
    \node[neuron] (N223) at (1.5+\nn,-1) {};
    \node[neuron] (N231) at (-0.5+\nn,-2) {};
    \node[neuron] (N232) at (0.5+\nn,-2) {};
    \node[neuron] (N233) at (1.5+\nn,-2) {};
    \node[neuron] (N241) at (0.5+\nn,-3) {};
    \node[neuron] (N251) at (-0.5+\nn,-4) {};
    \node[neuron] (N252) at (0.5+\nn,-4) {};
    \node[neuron] (N253) at (1.5+\nn,-4) {};
    \node[neuron] (N261) at (-0.5+\nn,-5) {};
    \node[neuron] (N262) at (0.5+\nn,-5) {};
    \node[neuron] (N263) at (1.5+\nn,-5) {};    
    \node[neuron] (N271) at (0.5+\nn,-6) {};      
    
    \draw[->] (N211) -- (N221);\draw[->] (N211) -- (N222);\draw[->] (N211) -- (N223);
    \draw[->] (N212) -- (N221);\draw[->] (N212) -- (N222);\draw[->] (N212) -- (N223);
    \draw[->] (N221) -- (N231);\draw[->] (N221) -- (N232);\draw[->] (N221) -- (N233);
    \draw[->] (N222) -- (N231);\draw[->] (N222) -- (N232);\draw[->] (N222) -- (N233);
    \draw[->] (N223) -- (N231);\draw[->] (N223) -- (N232);\draw[->] (N223) -- (N233); 
    \draw[->] (N231) -- (N241);\draw[->] (N232) -- (N241);\draw[->] (N233) -- (N241); 
    \draw[->] (N241) -- (N251);\draw[->] (N241) -- (N252);\draw[->] (N241) -- (N253); 
    \draw[->] (N251) -- (N261);\draw[->] (N251) -- (N262);\draw[->] (N251) -- (N263); 
    \draw[->] (N252) -- (N261);\draw[->] (N252) -- (N262);\draw[->] (N252) -- (N263); 
    \draw[->] (N253) -- (N261);\draw[->] (N253) -- (N262);\draw[->] (N253) -- (N263); 
    \draw[->] (N261) -- (N271);\draw[->] (N262) -- (N271);\draw[->] (N263) -- (N271);     
    
    \draw[dashed,->] (-1.25,-2.5) to[out=90, in=270] (-1.25,0.25) to[out=90, in=90] (N111);
    \draw[->] (N141) to[out=0, in=270] (2.25,-2.5) to[out=90, in=270] (2.25,0.25) to[out=90, in=90] (N211);
    \draw[dashed] (N241) to[out=0, in=270] (2.25+\nn,-2.5) to[out=90, in=270] (2.25+\nn,0.25);
    
    \draw[->] (N112)+(0,0.8) node[above]{$u_{t-1}$} -- (N112);
    \draw[->] (N212)+(0,0.8) node[above]{$u_t$} -- (N212);
    \draw (N141)+(1,0) node[below]{$\hat{x}_{1,t}$};
    \draw (N241)+(1,0) node[below]{$\hat{x}_{1,t+1}$};
    \draw (N111)+(-0.6,0.6) node[above]{$\hat{x}_{1,t-1}$};
    
    \draw[<-]  (N171)+(0,-0.8) node[below]{$\hat{y}_{1,t}$} -- (N171);
    \draw[<-]  (N271)+(0,-0.8) node[below]{$\hat{y}_{1,t+1}$} -- (N271);
    \end{tikzpicture}
  \caption{An example of a recurrent neural network. The figure illustrates the neural 
  network structure of residual $r_5$ in Eq.~\eqref{eq:r5} that is used to compute $\hat{y}_{t}$.}
  \label{fig:neural_network}
\end{figure}

\subsection{Model-Based Fault Diagnosis}

In model-based fault diagnosis, faults are identified by detecting inconsistencies 
between sensor data $y$ and predictions $\hat{y}$ from a physical-based model 
of the system, using residuals $r = y - \hat{y}$. To generate residuals require 
analytical redundancy in the model \cite{trave2014bridging}. A residual is a function 
of known variables and is, ideally, zero in the nominal case. Because of model 
uncertainties and sensor noise, different types of statistical tests are used to 
determine when a significant change in the residual output has occurred. Each residual 
models nominal system behavior, and can thus be interpreted as an anomaly 
classifier \cite{gupta2014outlier}.

\subsubsection{Structural Methods}

\begin{figure}[h]
      \begin{tikzpicture}
        \node[anchor=south west,inner sep=0] (image) at (0,0) {
          \resizebox{0.95\columnwidth}{!}{\includegraphics{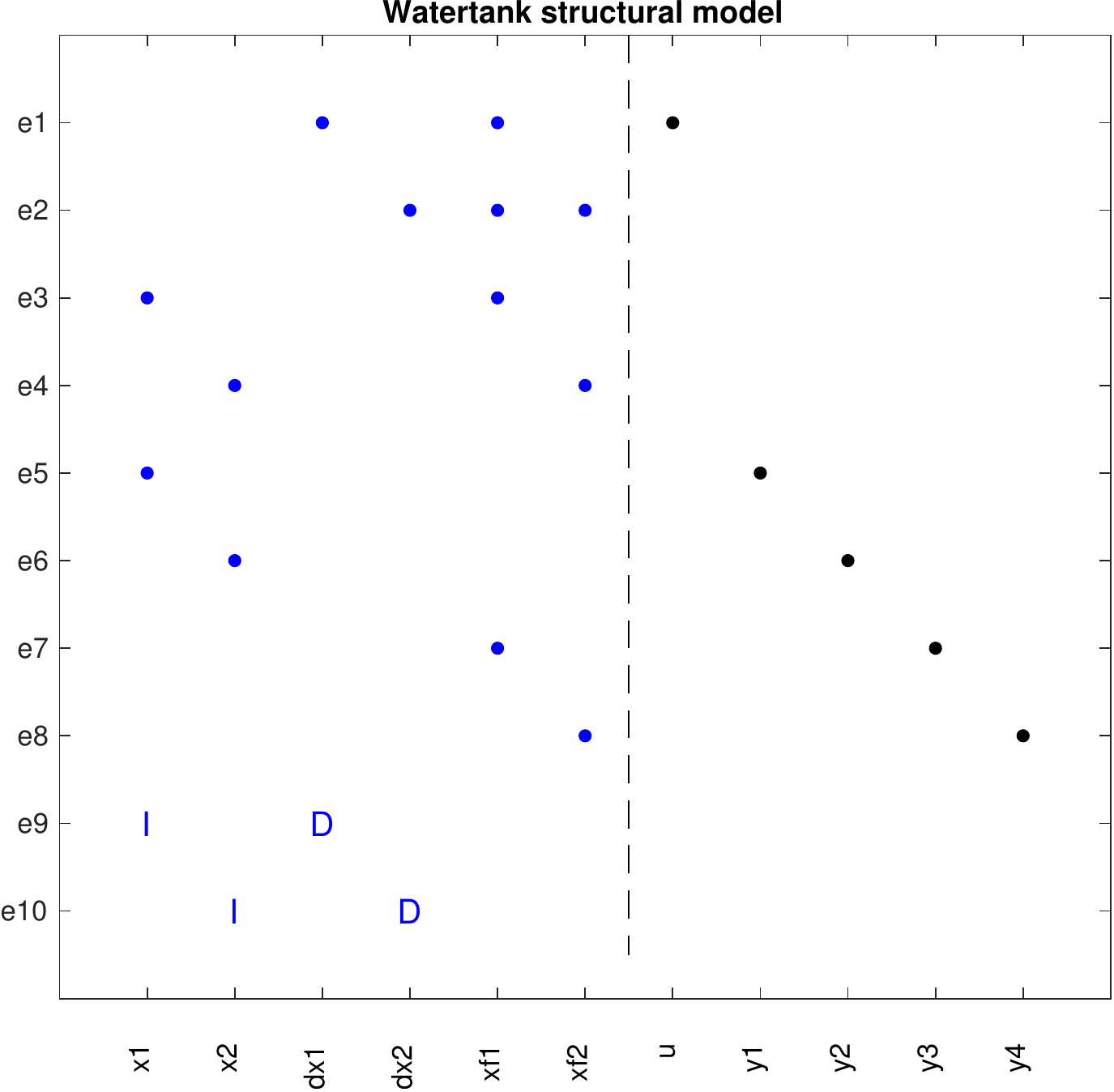}}
        };
        \begin{scope}[x={(image.south east)},y={(image.north west)}]
        \end{scope}
      \end{tikzpicture}
  \caption{A structural representation of the two tank model.}
  \label{fig:structural_model}
\end{figure}

A useful analysis tool for model-based diagnosis is structural methods \cite{blanke2006diagnosis}.  
A structural model is a bipartite graph describing the relationship between 
variables and equations and can be represented as an incidence matrix. 
Figure~\ref{fig:structural_model} shows an example of an structural representation 
of the two tank model Eq.~\eqref{eq:watertank_model}. Each row represents a model 
equation and each column a model variable. Equations $e_9$ and $e_{10}$ are 
used in the structural model to state the relationship that $\dot{x} = \frac{dx}{dt}$
where $I$ in Figure~\ref{fig:structural_model} is used in these equations to highlight 
the state variable and $D$ its derivative \cite{frisk2012diagnosability}. 

The structural model is 
not dependent on the actual analytical expression which makes it a useful tool for 
analysis during early system design since no parameter values are needed. By 
using a method called Dulmage-Mendelsohn decomposition on the structural model, 
it is possible to, for example, perform fault detectability and isolability analysis but also find 
redundant equation sets for residual generation \cite{krysander2008efficient}.

An example of a redundant equation set given Eq.~\eqref{eq:watertank_model} 
is $\{e_1, e_3, e_5\}$. The three equations contain two unknown variables 
$x_1$ and $x_{f,1}$ and can be used to generate a residual. Matching 
algorithms can be used with redundant equation sets to generate residuals, 
see for example \cite{frisk2017toolbox}. In principle, a matching algorithm finds 
a computational sequence describing how the known signals should be used 
to, sequentially, compute the unknown variables in the model using the 
equation set when one of the equations is used as a residual equation.
  
\begin{example} 
Asssume that the functions $h_1$ and $g_1$ in Eq.~\eqref{eq:watertank_model} 
are known. The equation set $\{e_1, e_3, e_5\}$ can be used to design 
different residuals, for example, 
\begin{equation}
\begin{aligned}
\dot{\hat{x}}_1 &= h_1 \left( g_1(\hat{x}_1), u \right) \\
r &= y_1 - \hat{x}_1 
\end{aligned}
\label{eq:integral_causality}
\end{equation}
or 
\begin{equation}
r = \dot{y}_1 - h_1 \left( g_1(\hat{x}_1), u \right). 
\label{eq:derivative_causality}
\end{equation}    
\end{example}
Equation \eqref{eq:integral_causality} is an example of a residual with integral causality 
and Eq.~\eqref{eq:derivative_causality} with derivative causality \cite{frisk2012diagnosability}.
If the redundant equation set does not contain any dynamic equations the residual is said 
to be an algebraic relation. In this work, only integral causality will be considered for 
residuals with dynamic equations.    
  
\subsubsection{Model-Based Residual Design}

Depending on which redundant set of model equations is used to generate a 
residual, the residual will be sensitive to faults in a certain part of the system. 
If different residuals are designed using different sets of redundant equations,
referred to as the \emph{model support} of the residual, the set of residuals 
will give a specific fault pattern depending on where a fault occurs in the system 
\cite{trave2014bridging}. Residuals that are designed to monitor the part 
of the system where a fault occurs should, ideally, deviate from their nominal 
behavior, while the other residuals should not be affected. By comparing the model 
support of the residuals that have deviated from their nominal behavior, it is 
possible to identify possible locations of the fault in the system based on 
the model equations \cite{pucel2009diagnosability}.

%% file: residual.tex
\section{Neural Network-Based Residual Generation}
Based on the structural model of the system, a set of different redundant 
equation sets are identified using the fault diagnosis toolbox \cite{frisk2017toolbox}. 
Based on each redundant equation set, a residual is modeled using a 
recurrent neural network where the location of the state variables are given 
by the structural model. For example, the residual in Eq.~\eqref{eq:integral_causality} 
is reformulated as 
\begin{equation}
\begin{aligned}
\dot{\hat{x}}_{1,t} &= \xi \left( \hat{x}_{1,t-1}, u_{t-1} \right) \\
r_{t} &= y_{1,t} - \hat{x}_{1,t} 
\end{aligned}
\end{equation} 
where subscript $t$ denotes time index, the unknown function 
$\xi: \mathbb{R}^2 \rightarrow \mathbb{R}$ is modeled using a neural 
network and the state variable is approximated using the Euler forward method to 
formulate a time-discrete model. The resulting residual function is a recurrent 
neural network with only one state variable, as illustrated in 
Figure~\ref{fig:neural_network}.  

Similarly, six additional residuals are implemented from different redundant 
equation sets, using the principles described in for example 
\cite{frisk2012diagnosability}, and the final residual set is summarized as follows:
\begin{align}
r_{1,t} &= y_{4,t} - \xi_1\left(y_{2,t}\right) \label{eq:r1}\\[2ex]
r_{2,t} &= y_{3,t} - \xi_2\left(y_{1,t}\right) \label{eq:r2}\\[2ex]
\hat{x}_{2,t} &= \xi_{3a}\left( y_{3,t-1}, \hat{x}_{2,t-1} \right) \nonumber\\
r_{3,t} &= y_{4,t} - \xi_{3b}\left(\hat{x}_{2,t}\right)  \label{eq:r3}\\[2ex]
\hat{x}_{2,t} &= \xi_{4}\left( y_{1,t-1}, \hat{x}_{2,t-1} \right) \nonumber\\
r_{4,t} &= y_{2,t} - \hat{x}_{2,t} \label{eq:r4}\\[2ex]
\hat{x}_{1,t} &= \xi_{5a}\left( \hat{x}_{1,t-1}, u_{t-1} \right) \nonumber\\
r_{5,t} &= y_{3,t} - \xi_{5b}\left(\hat{x}_{1,t}\right) \label{eq:r5}\\[2ex]
\hat{x}_{1,t} &= \hat{x}_{1,t-1} + \xi_{6}\left( y_{3,t-1}, u_{t-1} \right) + 0.01\left( y_{1,t} - \hat{x}_{1,t}\right) \nonumber\\
r_{6,t} &= y_{1,t} - \hat{x}_{1,t}  \label{eq:r6}\\[2ex]
\hat{x}_{1,t} &= \xi_{7a}\left( \hat{x}_{1,t-1}, u_{t-1} \right) \nonumber\\
\hat{x}_{2,t} &= \xi_{7b}\left( \hat{x}_{1,t-1}, \hat{x}_{2,t-1} \right) \nonumber\\
r_{7,t} &= y_{2,t} - \hat{x}_{2,t}  \label{eq:r7}
\end{align} 
where each function $\xi_i(\cdot)$ is modeled as a neural network. Residuals $r_3, ..., r_7$ are 
modeled as recurrent neural networks with similar structures as in Figure~\ref{fig:neural_network}. 
The model support of all residuals are summarized in Table~\ref{tab:model_support}.

\begin{table}
\centering
\caption{A summary of the equation sets used to design each neural network-based residual.}
\label{tab:model_support}
\begin{center}
    \begin{tabular}{l c c c c c c c c c}
     \hline
          	      & $e_1$ & $e_2$ & $e_3$ & $e_4$ & $e_5$ & $e_6$ & $e_7$ & $e_8$ \\
    \hline
   $r_1$           &            &            &            & X        &            & X         &            & X       \\
   $r_2$           &            &            &   X       &           &    X     &             &  X       &           \\
   $r_3$           &            &   X       &            &  X       &            &            &   X      &   X     \\
   $r_4$           &            &  X        &  X       &  X       &  X       &  X        &            &          \\
   $r_5$           &  X       &            &   X       &           &            &             &   X      &          \\
   $r_6$           &  X       &            &            &           &  X        &            &   X       &          \\
   $r_7$           & X        &   X      &    X     &   X      &             &  X        &            &          \\
    \hline
    \end{tabular}
\end{center}
\end{table}

Residuals $r_1$ and $r_2$ are static algebraic relations modeling the relation between 
measured water levels in each tank and the measured outflow in each tank. Residuals 
$r_3, ..., r_7$ have internal dynamics, where a small feedback term was introduced in 
the dynamic equation of residual $r_6$ in Eq.~\eqref{eq:r6} because of difficulties to achieve 
satisfactory prediction error when training the model. Also, note that in $r_6$, with 
respect to the other dynamic residuals, the term $\hat{x}_{1,t-1}$ is not included as an 
input in the neural network model $\xi_{6}(\cdot)$ but kept outside. This is necessary to 
maintain the redundant model structure and to make sure that faults not directly affecting 
the equations in the model support of $r_6$ are decoupled. Note that $r_6$ resembles a 
ResNet structure \cite{he2016deep}. 

Ideal fault localization performance, i.e. when model uncertainties are not considered, 
given the selected residual set is summarized in Table~\ref{tab:localization}. A fault in 
$e_i$ is isolable from a fault in $e_j$ if there is a residual where $e_i$ is part of its 
model support but not $e_j$. An X at position $(i,j)$ means that a fault affecting 
equation $e_i$ cannot be isolated from a fault affecting equation $e_j$. Based on 
the selected residuals it is possible to, ideally, localize a fault to a part of the system 
modeled by one equation. The exceptions are faults in $e_2$, $e_6$, and $e_8$ that 
cannot be isolated from a fault in $e_4$. 

\begin{table}
\centering
\caption{A summary of the equation sets used to design each neural network-based residual.}
\label{tab:localization}
\begin{center}
    \begin{tabular}{l c c c c c c c c c}
     \hline
          	      & $e_1$ & $e_2$ & $e_3$ & $e_4$ & $e_5$ & $e_6$ & $e_7$ & $e_8$ \\
    \hline
   $e_1$           &   X      &            &            &           &            &            &            &          \\
   $e_2$           &            &    X     &            &    X    &           &             &          &           \\
   $e_3$           &            &            &   X      &           &            &            &           &          \\
   $e_4$           &            &            &           &  X       &           &            &            &          \\
   $e_5$           &           &            &           &           &    X      &             &           &          \\
   $e_6$           &           &            &            &   X     &            &    X      &            &          \\
   $e_7$           &           &           &           &           &             &            &     X     &          \\
   $e_8$           &           &           &           &   X     &             &            &            &    X    \\
    \hline
    \end{tabular}
\end{center}
\end{table}

\subsection{Implementation and Training of Residuals}
Each residual is implemented in Python and PyTorch and trained using simulated fault-free 
data, see Figure~\ref{fig:sim_data}. Each non-linear function $\xi(\cdot)$ is here modeled 
using a neural network with three hidden layers, 32 neurons in each layer, and ReLU 
as activation function. The training is performed by simulating the system and minimizing 
the mean square error $\sum_t (y_t-\hat{y}_t)^2$ using the ADAM solver 
\cite{kingma2014adam} and truncated back-propagation through time 
\cite{werbos1990backpropagation}. It is important that training data are 
representative of nominal system operation since model validity is not 
expected for operating points not covered by training data.    

\begin{figure}[h]
      \begin{tikzpicture}
        \node[anchor=south west,inner sep=0] (image) at (0,0) {
          \resizebox{0.98\columnwidth}{!}{\includegraphics{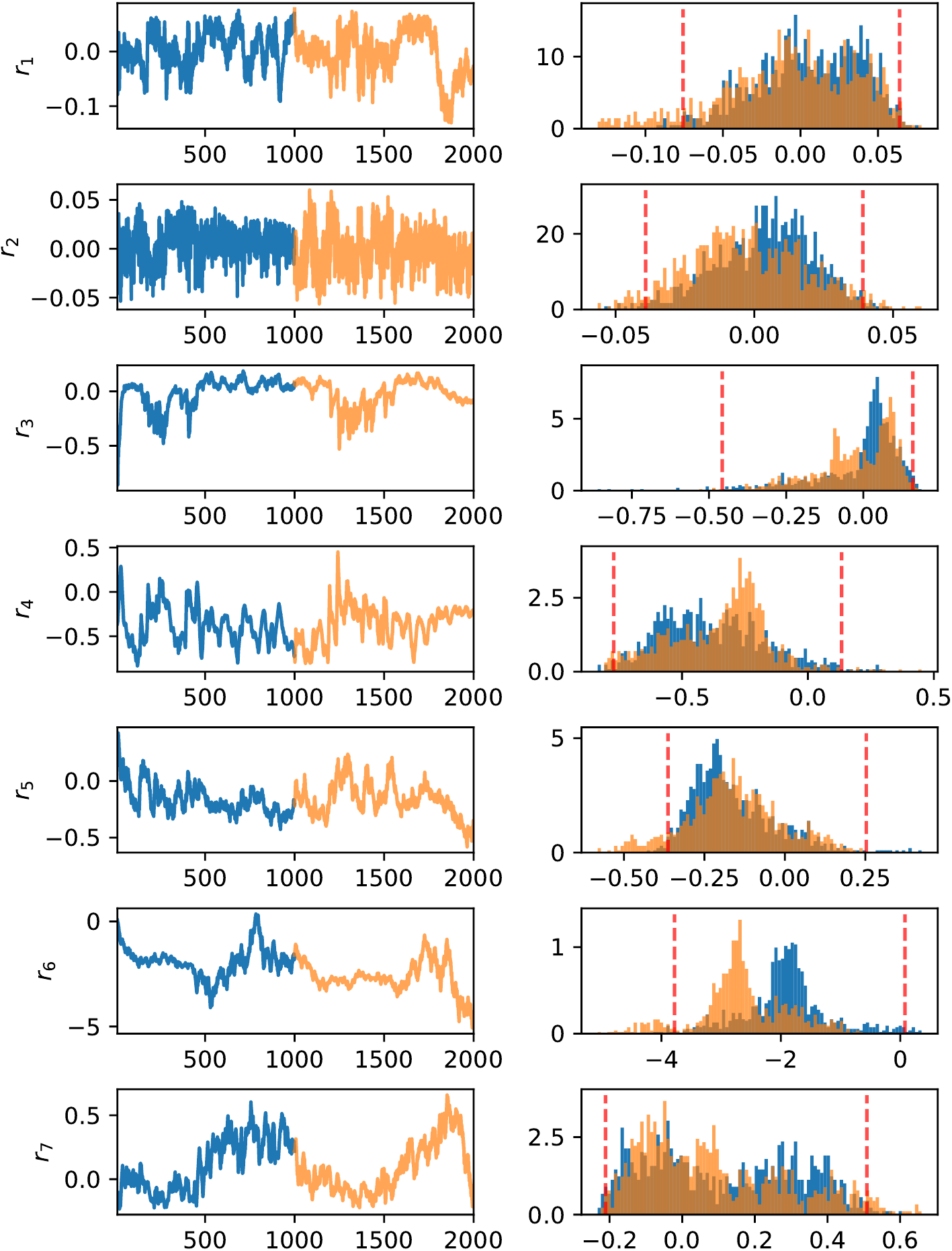}}
        };
        \begin{scope}[x={(image.south east)},y={(image.north west)}]
        \end{scope}
      \end{tikzpicture}
  \caption{The residual outputs using simulated fault-free data.}
  \label{fig:res_nf}
\end{figure}

An example of the evaluated residual outputs in the nominal case are shown in 
Figure~\ref{fig:res_nf}. The left column shows the residual time-series data, while the 
second column shows the histogram of each residual. The two histograms in each plot 
show the residual distribution during the first and second half of the time-series. The red 
dashed lines represent the $1\%$ and $99\%$ quantiles of the blue histograms 
representing the first half of the data set. These will be used analyze the residual 
outputs when the distributions are affected by different faults. Note that more 
sophisticated change detection algorithms can be used, instead of thresholding the 
residual, to automatically detect changes in the residual output, for example 
CUmulative SUM (CUSUM) \cite{page1954continuous}.  

%% file: evaluation.tex
\section{Evaluation}
To evaluate the fault localization performance of the neural network-based residuals, 
different fault scenarios are simulated. For single-fault scenarios, fault localization can 
be performed by analyzing the intersection of the model support of all residuals that 
significantly deviate from their nominal behavior. To handle multiple-fault scenarios, 
minimal hitting set algorithms, such as the one proposed in \cite{de1987diagnosing}, 
can be applied to identify likely fault localizations.      

\begin{figure}[h]
      \begin{tikzpicture}
        \node[anchor=south west,inner sep=0] (image) at (0,0) {
          \resizebox{0.98\columnwidth}{!}{\includegraphics{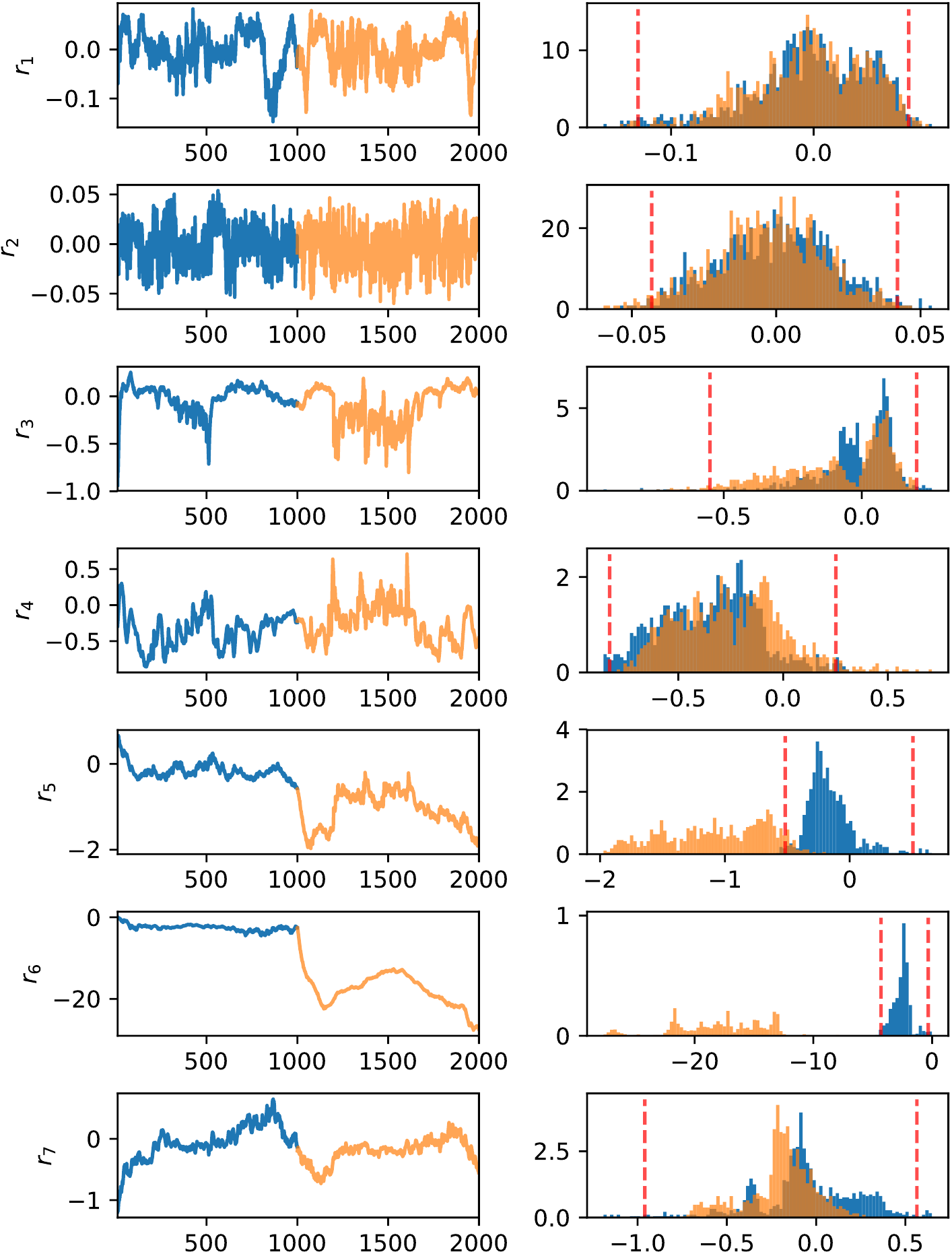}}
        };
        \begin{scope}[x={(image.south east)},y={(image.north west)}]
        \end{scope}
      \end{tikzpicture}
  \caption{The residual outputs when simulating a leakage in tank one.}
  \label{fig:res_f2}
\end{figure}

The first simulated fault scenario is a leakage fault in tank one occurring after 
sample 1000. The residual outputs are shown in Figure~\ref{fig:res_f2} where the time-series  
data are plotted in the left column and the histogram of the residuals, before and 
after the fault occurs, in the right column. When comparing the distributions in the 
right column it is visible that residuals $r_5$ and $r_6$ deviate significantly from 
their nominal behavior while the other ones does not. 

When comparing the model support for $r_5$ and $r_6$ in 
Table~\ref{tab:model_support}, the intersection of the corresponding equation sets 
is $\{e_1,e_7\}$, indicating that the fault should affect the part of the system 
described by one of the two equations. Equation $e_1$ describes the water level 
dynamics in tank one and $e_7$ the sensor measuring the outflow from tank one.
A leakage will affect the dynamics of the water level in the tank since there is an 
additional outflow from the tank not captured by the nominal model $e_1$ thus 
correctly narrowing down the location of the fault. Ideally, $r_7$ should also react 
to the leakage but is not deviating significantly in this case. An explanation could be 
that the accuracy of the residual model is not good enough to distinguish the fault. 

\begin{figure}[h]
      \begin{tikzpicture}
        \node[anchor=south west,inner sep=0] (image) at (0,0) {
          \resizebox{0.98\columnwidth}{!}{\includegraphics{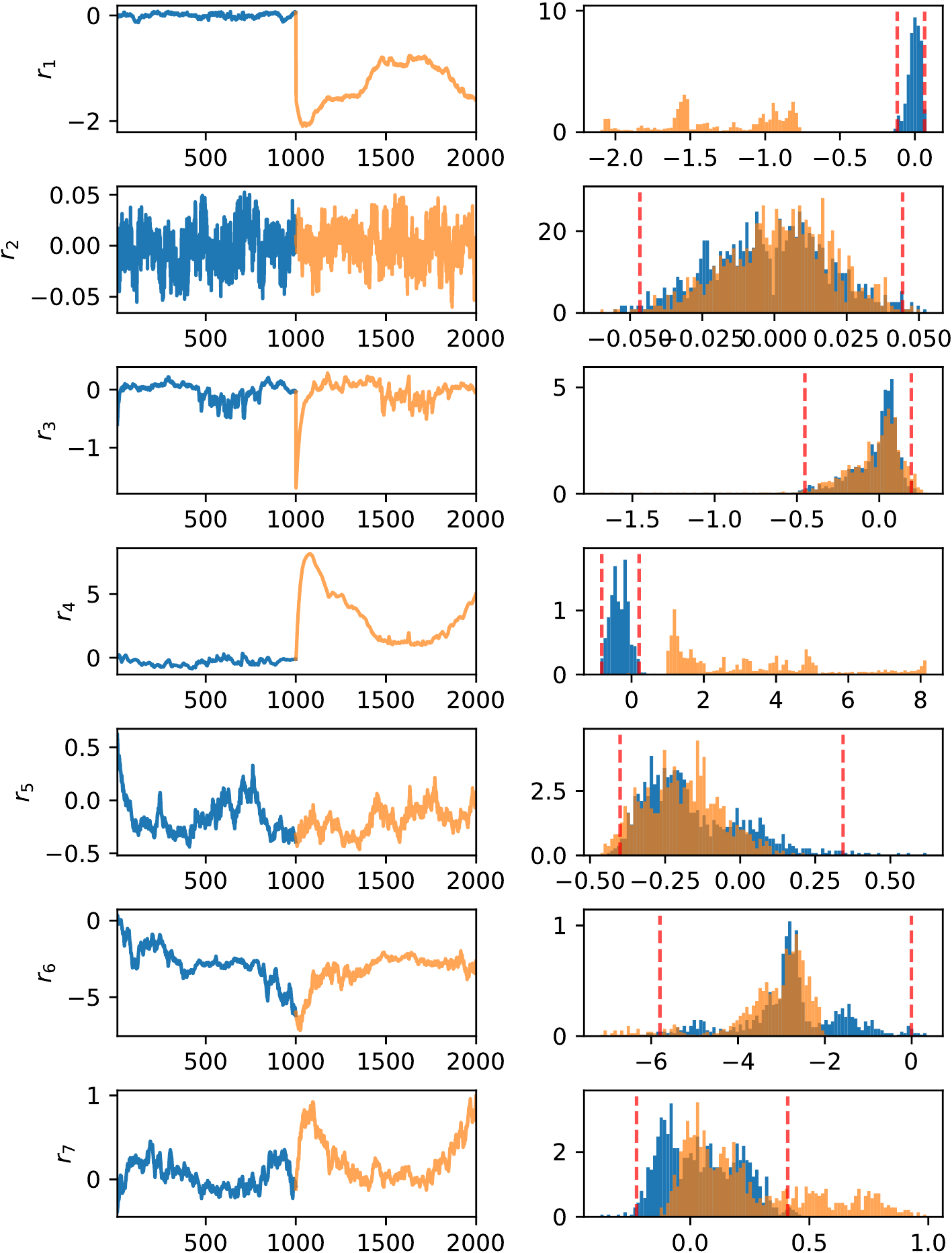}}
        };
        \begin{scope}[x={(image.south east)},y={(image.north west)}]
        \end{scope}
      \end{tikzpicture}
  \caption{The residual outputs when simulating clogging in the outflow pipe of tank two.}
  \label{fig:res_f5}
\end{figure}

In the second fault scenario, a clogging affecting the outflow from tank two is simulated 
and the residual outputs are shown in Figure~\ref{fig:res_f5}. In this case, there is 
a significant change in the distributions of residuals $r_1$ and $r_4$, while a 
small change can be noticed in $r_7$. Based on the model support in 
Table~\ref{tab:model_support} the intersection is $\{e_4,e_6\}$. Equation $e_4$ 
describes the relation between water level in tank one and the resulting outflow 
and $e_6$ the sensor measuring the level in tank two. The clogging fault is 
identified since the fault results in a decreased outflow described 
by $e_4$. Note that residual $r_3$, which is sensitive to a fault in $e_4$ 
makes a sudden change when the fault occurs but then goes back to nominal 
behavior. If a change detection algorithm applied to $r_3$ also triggers an alarm, 
$e_4$ would be isolated uniquely. 

The results from the two fault scenarios show that the trained set of neural 
network-based residuals can be used to identify the fault location in the actual 
system. Even if there were more than one equation where the fault could be 
located in the two scenarios, it gives useful information to a technician where 
to start troubleshooting. 

%% file: conclusions.tex
\section{Conclusions}

The case study shows that it is possible to perform fault localization 
of unknown faults using neural network-based residuals without the 
need of training data from faults. Developing accurate physical-based 
models can be time consuming. Hybrid methods combining 
qualitative models and machine learning can be one solution to reduce 
development time while still be make to make use of the structural 
properties of the physical-based model. 